\documentclass[11pt]{article}
\usepackage[a4paper, total={6in, 8in}]{geometry}
\usepackage{lipsum}
\usepackage{amssymb}

\usepackage{mathtools}
\usepackage{caption,comment,verbatim}
\usepackage{subcaption}
\usepackage{float}
\usepackage{amsmath}
\usepackage{lastpage}
\usepackage{fancyhdr}
\usepackage{url}

\usepackage{commath}
\usepackage{blindtext}

\title{Modelling surface waves on shear current with quadratic depth-dependence }

\author{Conor Curtin $^a$ and Rossen Ivanov $^b$  \thanks{member of the Institute for Advanced Physical Studies, 111 Tsarigradsko shose Blvd., Sofia 1784, Bulgaria}\\
\phantom{*}
\\ School of Mathematics and Statistics, \\Technological University Dublin, \\ City Campus, Grangegorman Lower, \\ Dublin D07 ADY7, Ireland\\
\phantom{8}\\
$^a$ Email: Conor.Curtin@tudublin.ie \\
$^b$ Email: Rossen.Ivanov@tudublin.ie}

\begin{document}

\maketitle

\begin{abstract}
    The currents in the ocean have a serious impact on ocean dynamics, since they affect the transport of mass and thus the distribution of salinity, nutrients and pollutants. In many physically important situations the current depends quadratic-ally on the depth. We consider a single layer of fluid and study the propagation of the surface waves in the presence of depth-dependent current with quadratic profile.
We select the scale of parameters and quantities, which are typical for the Boussinesq propagation regime (long wave and small amplitude limit) and we also derive the well known KdV model for the surface waves interacting with current.
    \end{abstract}

\section{Introduction}
Currents are steady mean flows of ocean water in a prevailing direction, while waves are periodic motion of water disturbances. For surface gravity waves on an inviscid fluid the total energy consists of the potential energy (resulting from the displacement of the mass of water from a position of equilibrium under the gravitational field) and the kinetic energy (due to the motion of the water particles throughout the fluid). The wave motion is related to oscillation of the total energy between these two forms, thus waves mainly transport energy, not matter. Currents on the other hand are substantially related to the mass transport, which involves mainly water and salt, but also pollutants and nutrients. The ocean dynamics is essentially nonlinear which leads to complicated interactions between the wave and current components of the velocity field. In the case of linear shear, the current brings only a constant vorticity (assuming two-dimensional simplified picture of the fluid motion). The constant vorticity automatically satisfies the vorticity equation and allows a lot of further simplifications, including modification of the Hamiltonian formulation, see for example the sequel of papers \cite{NearlyHamiltonian, Curtin,CIMT,IM,W,J} and also for internal waves and constant vorticity in each layer \cite{CoIv2,CIM-16,CuIv,Iv17}. There are many situations however, where the current has a non-linear depth profile. Here are a few examples:

(i) Equatorial Undercurrent (EUC). The EUC is one of the strongest currents in the ocean, discovered in 1952 by T. Cromwell, at depths between 100 m and 200 m. The underlying currents are usually wind-driven and confined to the near-surface. 3-4 m high gravity waves are common on the surface of the
ocean, while large internal waves (with heights in excess of 30 m) propagate as oscillations of the thermocline (the interface separating the two adjacent layers of different constant density).
In Stommel's model, \cite{Sto,Boyd} the following quadratic formula for the depth-dependent current $U(z)$ near the surface is derived:
\begin{equation*}
    U(z)=U(0)+\frac{\tau h}{2 \nu \rho} \left( \left( 1+\frac{z}{h}\right)^2 -1\right)
\end{equation*}
where $U(0)=(2/3)(\tau h/ 2 \nu \rho)$ is the current on the surface, $\tau $ is the wind stress, $\nu$ is the eddy viscosity (which is assumed constant near the surface), $\rho$ is the water density and $h$ is the depth of the ocean (at the top surface $z=0,$ at the bottom $z=-h$). A general derivation for a non-constant viscosity is presented in \cite{CJ}.

(ii) Another author, \cite{Mamaev} proposed a quartic polynomial formula for the EUC profile, which is valid throughout the whole depth and describes the flow reversal,
\begin{equation}
    U(z)=U(0)\left(1+k_1\frac{z}{h}+k_2\frac{z^2}{h^2}+k_3\frac{z^3}{h^3}+k_4\frac{z^4}{h^4}\right),
\end{equation} where $k_1,k_2,k_3$ and $k_4$ are related to some physical parameters of the ocean.
For small depths, $|z|\ll h,$ clearly, one can use only the quadratic approximation.

(iii) The current in a closed "rectangular" sea due to wind is given by the formula \cite{Mamaev}
\begin{equation}
    U(z)=U(0)\left(1+4\frac{z}{h}+3\frac{z^2}{h^2}    \right).
\end{equation}
This formula is valid for the whole depth and at the bottom $z=-h$ it gives $U(-h)=0.$

(iv) In \cite{Bow} a quadratic formula for the tidal current near the surface of the ocean
$(|z| < \alpha' h)$ where $\alpha' \approx 0.14$ is proposed:
\begin{equation*}
    U(z)=U_1+U_2\left(1-\frac{z^2}{h^2}\right)
\end{equation*} where $U_1, U_2$ are empirical constants.

The list of course could be extended further by other examples, which leads to the necessity to include quadratic terms in the expression of the underlying shear current. In this study we aim to show that in the framework of the Boussinesq approximation the quadratic current (through its parameters) can be accommodated in the derivation of the model equations. In particular, we will illustrate this with the derivation of the well known KdV equation \cite{KdV}.

\begin{figure}[!ht]
\centering
\includegraphics[width=0.8 \textwidth]{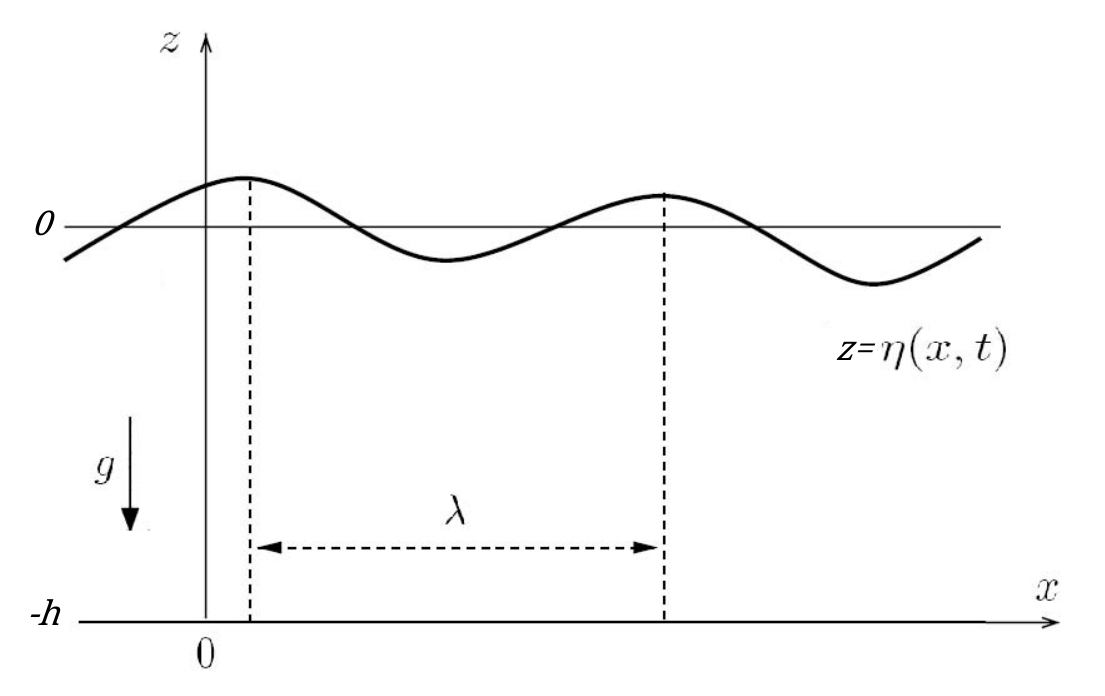}
 \caption{Coordinates and fluid domain.}\label{fig1}
\end{figure}

\section{Preliminaries- Setup and governing equations}\label{2.1}
We consider a two-dimensional motion of a single layer of fluid with free surface in the $(x,z)$ plane of a Cartesian coordinate system, as shown on Fig. \ref{fig1}. The unperturbed surface is at $z=0$ and the flat bed is at $z=-h,$ where $z$ is the vertical coordinate and $h>0.$ The wave elevation of the free surface is given by $z=\eta(x,t)$ where $\eta$ is a function of the horizontal variable $x$ and the time $t.$
Therefore we have
\begin{equation}\label{etaAverage}
\int_{\mathbb{R}} \eta(x,t) \, dx =0.
\end{equation}

The fluid motion is governed by Euler's equations
\begin{equation}\label{Eulereq}
	\begin{split}
u_t +u u_x + w u_z &=-\frac{1}{\rho} P_x  \\
w_t +u w_x +w w_z &=-\frac{1}{\rho} P_z -g
	\end{split}
\end{equation}
along with the incompressibility condition
\begin{equation}\label{div}
\text{div}\textbf{V}=u_x+w_z = 0 ,
\end{equation}
where $\textbf{V} (x,x,t)= (u(x,z,t),0,w(x,z,t))$ is the velocity field in the fluid body, $P(x,z,t)$ is the pressure, $\rho$ -- the constant density and $g$ -- the Earth's constant acceleration.
The systems \eqref{Eulereq} and \eqref{div} are complemented by the boundary conditions
\begin{equation}\label{BKS}
w=\eta_t + u \eta_x\quad{\rm on}\quad z=\eta(x,t),
\end{equation}
\begin{equation}
    w=0\quad{\rm on}\quad z=-h,
\end{equation}
and
\begin{equation}\label{P}
    P=P_{\rm atm} \, \, \text{on} \, \, z=\eta(x,t),
\end{equation}
were $P_{\rm atm} $ is the constant atmospheric pressure on the surface. The irrotational fluid motion in one layer has been heavily studied by now, see for example \cite{CGS,Lan} and the references therein. The main aspects of the wave-current interactions are presented in the monograph \cite{Constantin_2011}.

According to the Helmholtz theorem any vector field could be  resolved into the sum of an irrotational (curl-free) vector field and a solenoidal (divergence-free) vector field.  We assume in the usual manner, in line with Helmholtz's Theorem, that the velocity field decomposes into a potential part and a shear current $ (U(z),0) ;$ the former being irrotational and the latter will be $\textbf{quadratic}$ function of $z:$
\begin{equation}
    U(z)=\kappa +\gamma z +\beta z^2.
\end{equation} The divergence of $(U(z), 0)$ is obviously equal to zero.
Introducing a velocity potential $\varphi(x,z,t),$ and stream function $\psi(x,z,t),$ the velocity field is therefore given by
\begin{equation}\label{uv}
\begin{split}
u &= \varphi_x +U(z)=\varphi_x+\kappa +\gamma z +\beta z^2 = \psi_z \\
w &= \varphi_z = -\psi_x
\end{split}
\end{equation} where $\kappa = U(0), $   $\gamma$ and $\beta$ are constants, as required by the chosen model.
Moreover, due to \eqref{div} and \eqref{uv}
\begin{equation}
    \Delta \varphi(x,z,t) \equiv (\partial_x^2+\partial_z^2)\varphi(x,z,t)=0.
\end{equation}
Note that the vorticity is now $z-$ dependent. Indeed, it could be verified that the vorticity is $u_z-w_x= U'(z) =\gamma+2\beta z.$ Thus, the vorticity equation has to be taken into account.

The kinematic boundary condition on the surface \eqref{BKS} can be written also as
\begin{align}\label{etaphi}
\eta_t=(\varphi_z)_s-\eta_x\left((\varphi_x)_s +U(\eta)\right).
\end{align} where sub-index "s" means evaluation on the surface.
The boundary condition at the bottom has an equivalent form via the velocity potential:
$$\varphi_z (x,-h)=0.$$

We must make some assumptions about the growth rate of our functions. We assume that the functions $\eta(x, t),$ and ${\varphi}(x, z, t)$ belong to the Schwartz space of functions $\mathcal{S}(\mathbb{R})$ with respect to the $x$ variable (for any $z$ and $t$). This reflects the localised nature of the wave disturbances.
This assumption implies also the following  \begin{equation} \label{vanish}
\lim_{|x|\rightarrow \infty}\eta(x,t)=0, \quad \lim_{|x|\rightarrow \infty}{ {\varphi}}(x,z,t)=0.
\end{equation}

The velocity field \eqref{uv} can be extended from its boundary values to the volume of the fluid by analyticity. Indeed, the function $\Psi=\psi-\kappa z -(1/2) \gamma z^2-(1/3)\beta z^3,$  together with $\varphi$ satisfy the Cauchy-Riemann equations, $\varphi_x=\Psi_z,$  $\varphi_z=-\Psi_x.$ Thus, $F=\varphi(x,z)+i\Psi(x,z)$ is an analytic function in the fluid domain, which is determined by its values on the boundary - that is, the surface and the bottom. See more details in \cite{Delia}.

\section{Bernoulli's equation}

We now proceed to rewrite the governing equations in terms of the functions $\varphi$ and $\psi,$
\begin{equation}\label{A}
\begin{split}
\varphi_{xt} +\left(\varphi_x +U\right)\varphi_{xx} + \varphi_z\left[\varphi_{xz} + U'(z)\right] = -\left(\frac{P}{\rho}\right)_x ,\\ \\
\varphi_{zt}+\left(\varphi_x +U\right) \varphi_{xz} +\varphi_z\varphi_{zz} = -\left(\frac{P}{\rho}\right)_z -g.
\end{split}
\end{equation}
Considering the first expression in \eqref{A} we may write
\begin{equation*}
\left[ \varphi_t +\frac{1}{2}\left(\varphi_x^2 +\varphi_z^2\right) + \frac{P}{\rho}\right]_x + U(z)\varphi_{xx} + \varphi_z U'(z)=0,
\end{equation*}
or, using the "nabla" notation, $\abs{\nabla\varphi}^2 = \varphi_x^2 +\varphi_z^2 ,$
\begin{equation}\label{B}
\left[ \varphi_t +\frac{1}{2}\abs{\nabla \varphi}^2  + \frac{P}{\rho} +gz + U(z) \varphi_x \right]_x - \psi_x U'(z) = 0 .
\end{equation}
Hence we may express \eqref{B} as a total derivative with respect to $x$
\begin{equation}
\left[\varphi_t +\frac{1}{2}\abs{\nabla \varphi}^2 + \frac{P}{\rho} +gz + U(z) \varphi_x  - \psi U'(z)\right]_x= 0.
\end{equation}
Integration with respect to $x$ gives a function of $z$ and $t.$ However, in what follows, for the sake of simplicity we are not going to highlight explicitly the $t-$ dependence.  Thus we obtain the equation
\begin{equation}\label{BB}
\begin{split}
\varphi_t +\frac{1}{2}\abs{\nabla \varphi}^2 + \frac{P}{\rho} +gz + U(z) \varphi_x  - \psi U'(z) &= \tilde{F}(z) , \\
\varphi_t +\frac{1}{2}\left[(\varphi_x + U)^2 +\varphi_z^2\right] +\frac{P}{\rho} +gz -U'(z) \psi &= \tilde{F}(z) + \frac{U^2}{2} = F(z).
\end{split}
\end{equation}
This is a way of writing the Bernoulli equation. We need to check however if this form is compatible to the second Euler equation. Transforming the second expression in \eqref{A} we obtain in a similar fashion
\begin{equation} \label{C}
\begin{split}
\left[\varphi_t +\frac{1}{2} \abs{\nabla \varphi}^2 +gz +\frac{P}{\rho}\right]_z +U(z) \varphi_{xz} &= 0, \\
\left[\varphi_t +\frac{1}{2} \abs{\nabla \varphi}^2 +gz +\frac{P}{\rho} +U \varphi_x \right]_z -U'(z) \varphi_x &=0.
\end{split}
\end{equation}
Using the expression $\varphi_x = \psi_z - U(z), $ we multiply both sides by $-U'(z)$ to obtain
\begin{equation}\label{D}
 -U '(z)\varphi_x = -U'(z) \left( \psi_z - U(z) \right)  = \left(\frac{U^2}{2}\right)_z -\left( U'(z) \psi\right)_z + U''(z) \psi
\end{equation}
Hence we may rewrite \eqref{C} using \eqref{D} as
\begin{equation*}
\left[ \varphi_t +\frac{1}{2} \abs{\nabla \varphi}^2 +gz +\frac{P}{\rho} +U\varphi_x +\frac{U^2}{2} -U'(z)\psi \right]_z + U''(z)\psi = 0
\end{equation*}
or, expanding  the gradient term in a similar fashion as in \eqref{BB}
\begin{equation}\label{CC}
\left[\varphi_t +\frac{1}{2}\left[(\varphi_x+U)^2 + \varphi_z^2\right] +\frac{P}{\rho} +gz -U'(z) \psi \right]_z +U''(z) \psi (x,z,t)= 0.
\end{equation}
Now, from \eqref{BB} and \eqref{CC} we have therefore
\begin{equation}\label{eqF}
    F'(z)+U''(z) \psi (x,z,t)= 0.
\end{equation}
This equation is actually the vorticity equation in the case of time-independent vorticity. In what follows we need the Bernoulli-type equation for the surface variables, $z=\eta(x,t)$ and therefore we need to solve for $F$
\begin{equation}\label{eqF-eta}
    F'(\eta)+2\beta  \psi (x,\eta,t)= 0.
\end{equation}

\section{Physical scales and approximations.}

In the case of fluid motion in one layer, the widely used non-dimensional scale parameters are $\varepsilon= \frac{\eta_{\text{max}}}{h}$ where $\eta_{\text{max}}$ is the maximal deviation of $\eta(x,t)$ from its equilibrium at $\eta=0,$ this could be either the maximal elevation or the maximal descent; $\delta=\frac{h}{\lambda},$ where $\lambda$ is the typical wavelength of the waves under study. We assume that both scale factors are small, $\varepsilon\ll 1$ requires consideration of small wave amplitudes, while $\delta \ll 1$ is the assumption for large wavelengths (or small depths). In addition, a crucial further assumption in our study is that $\varepsilon$ and $\delta^2$ are of the same order - this is the so-called Boussinesq scaling regime.

The governing equations could be written in terms of non-dimensional variables. This is possible, since all necessary dimensional combinations could be produced by the dimensional constants $g,\, h$ and $\rho.$ Moreover, then, the non-dimensional counterparts of $g,\, h$ and $\rho$ are clearly $1, \,1$ and $1,$  see for example \cite{CJ2}.
Thus, the governing equations will depend only on the non-dimensional variables and the scale-parameter
$\varepsilon.$  For example, the order of $\eta$ is the same as the order of $\varepsilon h$ or, in non-dimensional terms, just $\varepsilon.$ To keep track of the scales, we introduce the rule that the quantities with tilde are of order 1. For example instead of $\eta$ we write $\eta=\varepsilon\tilde{\eta},$ and we remember that $\tilde{\eta}$ is of order 1.

In what follows we are not going to perform explicitly the non-dimensionalisation, because we want to keep track of the constants mentioned above, but rather, we accept that $g=\Tilde{g},\, h=\Tilde{h}$ and $\rho$ are quantities of order 1, then $\eta $ is clearly of ordrer $\varepsilon,$ the $x-$derivative, acting on a monochromatic wave $\exp(ikx-i\omega t)$ has an eigenvalue $ik,$ where $k=\frac{2\pi}{\lambda}$ is the wave-number. Then $kh=2\pi \delta \sim \delta$ and we conclude that $\partial_x = \delta \partial_{\Tilde{x}}=\sqrt{\varepsilon}\partial_{\Tilde{x}}.$ We also assume that the constants $\kappa, \gamma, \beta$ are of order 1 (that is, their non-dimensional counterparts are of order 1). In the KdV regime also $\varphi_x $ is of order $ \varepsilon$ thus $\varphi = \sqrt{\varepsilon} \Tilde{\varphi}.$ There are of course other possible regimes, see for example the comments in \cite{Hor}.

\section{The Dirichlet - Neumann Operator and the Boussinesq regime}

The Bernoulli equation, together with \eqref{etaphi} describe completely the dynamics on the surface in terms of the variables $\eta(x,t)$ and the velocity potential evaluated on the surface $z=\eta(x,t),$ that is
$$ \xi(x,t)=  \varphi \left(x, \eta(x,t), t \right) =  (\varphi)_{s},$$ which belongs to $\mathcal{S}(\mathbb{R})$ with respect to the $x$ variable. Moreover,
\begin{equation}\label{xi_x}
    \xi_x=(\varphi_x)_s+(\varphi_z)_s \eta_x.
\end{equation}
\begin{equation}\label{xi_t}
    \xi_t=(\varphi_t)_s+(\varphi_z)_s \eta_t.
\end{equation}


In what follows we need the definition of the so-called Dirichlet - Neumann Operator (DNO), $G(\eta)$.
The DNO is defined as follows, see for example \cite{CraigGroves1,CGS}:
\begin{align} G(\eta)\xi= \left(\frac{\partial \varphi}{\partial {\bf n}} \right)_s \cdot \sqrt{1+\eta_x^2},
\end{align}
where $${\bf n}=\frac{(-\eta_x,1)}{\sqrt{1+\eta_x^2}}$$ is the outward unit normal to the fluid surface, thus
 \begin{equation}\label{dno1}
 \left(\frac{\partial \varphi}{\partial {\bf n}} \right)_s ={\bf n}\cdot (\nabla  \varphi)_s,
 \end{equation}
is the corresponding directional derivative, or equivalently,
\begin{equation} \label{Gxi}
G(\eta) \xi = -\eta_x (\varphi_x)_s +(\varphi_z)_s.\end{equation}
Then, from \eqref{etaphi}
\begin{align}\label{etat}
\eta_t =G(\eta)\xi- (\kappa+\gamma\eta + \beta \eta^2)\eta_x .
\end{align}
The DNO $G(\eta)$ can be expressed as an infinite asymptotic series in $\varepsilon.$ For this we refer for example to  \cite{CraigGroves1,Craig1993,Lan},
\begin{equation}\label{dnoG}
\begin{split}
    G(\eta)&= -(\delta^2) h\partial_{\Tilde{x}}^2-(\delta^4)\frac{1}{3} h^3 \partial_{\Tilde{x}} ^4 -\delta^2\varepsilon \partial_{\Tilde{x}} \Tilde{\eta} \partial_{\Tilde{x}}+ \mathcal{O}(\varepsilon^3),\\
    &= -\varepsilon h\partial_{\Tilde{x}}^2-\varepsilon^2 \frac{1}{3} h^3 \partial_{\Tilde{x}} ^4 -\varepsilon^2 \partial_{\Tilde{x}} \Tilde{\eta} \partial_{\Tilde{x}}  +\mathcal{O}(\varepsilon^3).
    \end{split}
\end{equation}
From \eqref{dnoG} and \eqref{etat} with scales we have
\begin{align}
       (\delta \varepsilon)  \Tilde{\eta}_{\Tilde{t}} &= [-\varepsilon h\partial_{\Tilde{x}}^2-\varepsilon^2 \frac{1}{3} h^3 \partial_{\Tilde{x}} ^4 -\varepsilon^2 \partial_{\Tilde{x}} \Tilde{\eta} \partial_{\Tilde{x}}]((\delta) \Tilde{\xi})- (\kappa+\varepsilon \gamma \Tilde{\eta }+ \varepsilon^2\beta \Tilde{\eta}^2)(\delta \varepsilon) \Tilde{\eta}_{\Tilde{x}} + \ldots, \nonumber \\
   \Tilde{\eta}_{\Tilde{t}} &=-h \Tilde{\xi}_{\Tilde{x}\Tilde{x}}-\varepsilon \frac{h^3}{3} \xi_{\Tilde{x}\Tilde{x}\Tilde{x}\Tilde{x}}-\varepsilon(\Tilde{\eta} \Tilde{\xi} _{\Tilde{x}})_{\Tilde{x}}-\kappa \Tilde{\eta}_{\Tilde{x}}
   -\varepsilon \gamma \Tilde{\eta} \Tilde{ \eta} _{\Tilde{x}} + \mathcal{O}(\varepsilon^2). \label{Eq4eta}
    \end{align}

The dynamics therefore is determined by \eqref{Eq4eta} and \eqref{BB}. We notice that \eqref{Eq4eta} in this approximation does not depend on $\beta.$ We need now to express the Bernoulli equation \eqref{BB} in terms of the variables $\eta$ and $\xi$ for the relevant scales.

The stream function on the surface is $\psi_s(x,t)=\psi(x, \eta(x,t),t)$ and therefore
\begin{equation}
    \frac{d \psi_s}{dx}= (\psi_x)_s+(\psi_z)_s \eta_x= (-w + u \eta_x)_s=-\eta_t
\end{equation} due to \eqref{BKS}. This gives
\begin{equation}
    \psi_s=-\int_{-\infty} ^x \eta_t(x',t) \, dx' \equiv -\partial_x^{-1}\eta_t.
\end{equation}
With \eqref{etat} we obtain further
\begin{equation}\label{psi_s_gen}
    \psi_s= -\partial_x^{-1}\eta_t= -\partial_x^{-1} [G(\eta)\xi]+ \kappa\eta +\frac{1}{2}\gamma\eta^2 + \frac{1}{3}\beta \eta^3  .
\end{equation}

\noindent Using \eqref{dnoG} in \eqref{psi_s_gen} we obtain
\begin{equation}\label{psi_s}
    \psi_s=\varepsilon (\kappa \Tilde{\eta}+h \Tilde{\xi}_{\Tilde{x}})+ \varepsilon^2\left(\frac{h^3}{3}\Tilde{\xi}_{\Tilde{x}\Tilde{x}\Tilde{x}}+\Tilde{\eta} \Tilde{\xi}_{\Tilde{x}}+\frac{\gamma}{2} \Tilde{\eta}^2   \right) + \mathcal{O}(\varepsilon^3).
\end{equation}
Equation \eqref{BB} on the surface $z=\eta (x,t)$ for the purposes of the Boussinesq regime requires $F(\eta)$ up to quantities of order $\varepsilon^2.$ Let us suppose that
\begin{equation}
  \Tilde{\eta}_{\Tilde{t}} =-(\kappa+c_0) \Tilde{\eta}_{\Tilde{x}} +    \mathcal{O}(\varepsilon ), \qquad
 \partial^{-1}_{\Tilde{x}}  \Tilde{\eta}_{\Tilde{t}} =-(\kappa+c_0) \Tilde{\eta} +    \mathcal{O}(\varepsilon),
\end{equation} for some constant $c_0,$ which represents the propagation wave speed in leading order. This constant will be determined later. Then from
\eqref{eqF-eta}, \eqref{psi_s_gen} and \eqref{etaAverage}  we have
\begin{equation}\label{eqF-eta1}
    F'(\eta)=-2\beta  \psi (x,\eta,t)= 2\beta \partial_x^{-1}\eta_t, \qquad F'(0)=0 ;
\end{equation} hence,
\begin{align}\label{eqF-eta2}
    F'(\eta)&=  -2\beta (\kappa+c_0) \varepsilon  \Tilde{\eta }   +\mathcal{O}(\varepsilon^2 )= -2\beta (\kappa+c_0) {\eta }  + \mathcal{O}(\varepsilon^2 ) , \\
     F(\eta)&= F_0(t)- \beta (\kappa+c_0) \eta^2 +    \mathcal{O}(\varepsilon^3),
\end{align} where $F_0(t)$ is a yet unknown function as a result of the integration.

In order to find $(\varphi_x)_s$ and $(\varphi_x)_z$ we solve \eqref{xi_x} and \eqref{Gxi} as a system of simultaneous linear equations, which gives
\begin{align}\label{phi_syst}
\begin{split}
(\varphi_x)_s\  & =\ \frac{1}{1+\eta_x^2}\ \Big( \xi_x - \eta_x G \xi \Big), \\
(\varphi_z)_s\   &=\  \frac{1}{1+\eta_x^2}\ \Big( G\xi  + \eta_x \xi_x\Big).
\end{split}
\end{align}
From \eqref{xi_t} we have
\begin{equation}
\label{phi_t}
(\varphi_t)_s=\xi_t-(\varphi_z)_s \eta_t.
\end{equation}
Now we have all of the ingredients to obtain the Bernoulli equation \eqref{BB} in terms of $\eta$ and $\xi.$
From the conditions \eqref{vanish} we determine
\begin{equation}
    \frac{P_{\text{atm}}}{\rho}+ \frac{\kappa^2}{2}=F_0
\end{equation} (that is, $F_0$ is a constant) and with \eqref{phi_t}
\begin{equation}\label{BB1}
    \xi_t - (\varphi_z)_s \eta_t + \frac{1}{2}[(\varphi_x)^2_s+(\varphi_z)^2_s] +(\varphi_x)_s U(\eta) + \frac{U^2(\eta)-\kappa^2}{2}
    -(\gamma+2 \beta \eta) \psi_s+\beta(\kappa+c_0)\eta^2=0.
\end{equation}
All quantities in \eqref{BB1} and their asymptotic expansions are known in terms of $\eta$ and $\xi,$  see for example
 \eqref{Eq4eta}, \eqref{phi_syst}, \eqref{psi_s} as well as the DNO expansion \eqref{dnoG}.

The asymptotic expansion could be obtain by introducing the scale factors and keeping in mind that only the terms of orders $\varepsilon$ and $\varepsilon^2$ should be kept. Note that
\begin{align*}
    \frac{1}{1+\eta_x^2}&=1-\varepsilon^2 \delta^2\Tilde{\eta}_{\Tilde{x}}^2+ \ldots =1+\mathcal{O}(\varepsilon^3), \\
    G \xi    & = -h \delta^3 \Tilde{ \xi}_{\Tilde{x}\Tilde{x}}+\ldots = \mathcal{O}(\delta^3)=\mathcal{O}(\varepsilon^{3/2})
\end{align*} and therefore the terms $2\eta_x \xi_x G\xi$ and $  (G\xi)^2$ from $(\varphi_x)^2_s+(\varphi_z)^2_s$  do not contribute. With the scale factors, the final form of \eqref{BB1} is
\begin{equation}\label{Eq4xi}
    \Tilde{\xi}_{\Tilde{t}} + (\kappa - \gamma h) \Tilde{\xi}_{\Tilde{x}} + g \Tilde{\eta} + \varepsilon \left(  \frac{1}{2} \Tilde{\xi}_{\Tilde{x}}^2 -\frac{\gamma h^3}{3}\Tilde{ \xi}_{\Tilde{x}\Tilde{x}\Tilde{x}}
    - 2 \beta h \Tilde{\eta} \Tilde{ \xi}_{\Tilde{x}} + \beta c_0 \Tilde{\eta}^2 \right)= \mathcal{O}(\varepsilon^2).
\end{equation}
The equations \eqref{Eq4xi} and \eqref{Eq4eta} are the equations in the Boussinesq approximation. We observe that \eqref{Eq4xi} depends on the parameter $\beta,$ which is the coefficient of the quadratic term of $U(z).$
Since the evolution equations in their asymptotic form are already in place, we drop all tildes from now on.
The parameter $\kappa=U(0)$ can be eliminated, if the equations are written in a reference frame that moves with speed $\kappa.$ Then a Galilean change $x \rightarrow x-\kappa t$ gives $\partial_t + \kappa \partial_x \rightarrow \partial_t.$  It is also convenient to use the variable $\mathfrak{u}:=\xi_x$ which has a meaning of tangent velocity on the surface. Then the Boussinesq-type equations become
\begin{align}
     &\mathfrak{u}_t - \gamma h \mathfrak{u}_x + g \eta _x+ \varepsilon \left(  \frac{1}{2} \mathfrak{u}^2 -\frac{\gamma h^3}{3} \mathfrak{u}_{xx}- 2 \beta h \eta \mathfrak{u} +\beta c_0 \eta^2 \right)_x= \mathcal{O}(\varepsilon^2). \label{eq4u1}\\
    & \eta_t +h \mathfrak{u}_{x}+\varepsilon \left(\frac{h^3}{3} \mathfrak{u}_{xxx}+(\eta \mathfrak{u})_x
   + \gamma \eta \eta_x \right)= \mathcal{O}(\varepsilon^2). \label{eq4eta1}
\end{align}
We kept the coefficient $\kappa$ until now in order to trace any couplings between the quadratic term, related to vorticity and the velocity on the surface.

The surface dynamics in this approximation has a quasi-Hamiltonian formulation (like the constant vorticity case \cite{NearlyHamiltonian} where $\gamma \ne 0,$ $\beta=0):$
\begin{align}
    \eta_t&= -\left( \frac{\delta H}{\delta \mathfrak{u}}\right)_x  \label{H1}\\
    \mathfrak{u}_t&= -\left( \frac{\delta H}{\delta \eta}\right)_x -\gamma \eta_t - \varepsilon 2\beta \eta_x \partial_x ^{-1}\eta_t=  -\left( \frac{\delta H}{\delta \eta}\right)_x +[\gamma \partial_x + \varepsilon 2\beta \eta_x ]  \left( \frac{\delta H}{\delta \mathfrak{u}}\right), \label{H2}
\end{align}
where \begin{align}
    H&= \frac{\varepsilon^2}{2}\left[\int_{\mathbb{R}} (h\mathfrak{u}^2+ g \eta^2)\, dx \right] +\frac{\varepsilon^3}{2}\left[\int_{\mathbb{R}} \left(-\frac{h^3}{3}\mathfrak{u}_x^2+ \frac{\gamma^2}{3} \eta^3 + \eta \mathfrak{u}^2+\gamma \eta^2 \mathfrak{u}\right)\, dx \right] +\mathcal{O}(\varepsilon^4)
\end{align} is the energy of the surface wave motion in terms of $\eta, \mathfrak{u}.$ The equation \eqref{H1} gives immediately \eqref{eq4eta1}, the other equation, \eqref{H2} with some straightforward simplification with \eqref{eq4eta1} leads to \eqref{eq4u1}.

We note that due to the equation $\eta_t=-c_0 \eta_x+....$ in the leading order approximation (see Section \ref{kdvsec}), equation \eqref{eq4u1} could be written in a non-evolutionary form (since $c_0$ is so far not explicitly determined) as
\begin{equation}\label{nonevol}
    \mathfrak{u}_t - \gamma h \mathfrak{u}_x + g \eta _x+ \varepsilon \left(  \frac{1}{2} \mathfrak{u}^2 -\frac{\gamma h^3}{3} \mathfrak{u}_{xx}- 2 \beta h \eta \mathfrak{u} \right)_x - \varepsilon 2\beta \eta \eta_t= \mathcal{O}(\varepsilon^2).
\end{equation}
 The $\beta-$related addition in \eqref{H2} is $\varepsilon 2\beta \eta_x \partial_x ^{-1}\eta_t$ and due to \eqref{eq4eta1} is
 \begin{align}
   \varepsilon 2\beta \eta_x \partial_x ^{-1}\eta_t=&\varepsilon 2\beta \eta_x \partial_x ^{-1}[-h \mathfrak{u}_x]+\mathcal{O}(\varepsilon^2)=
   -\varepsilon 2\beta h  \eta_x \mathfrak{u}+\mathcal{O}(\varepsilon^2) \nonumber\\
   =&-\varepsilon 2\beta h ( \eta \mathfrak{u})_x-\varepsilon 2\beta   \eta (-h\mathfrak{u}_x)+\mathcal{O}(\varepsilon^2) =-\varepsilon 2\beta h ( \eta \mathfrak{u})_x-\varepsilon 2\beta   \eta \eta_t +\mathcal{O}(\varepsilon^2),
 \end{align} which is exactly the $\beta-$related term in \eqref{nonevol}. The equation \eqref{eq4u1} therefore has also the following asymptotically equivalent form:
 \begin{equation}
     \mathfrak{u}_t - \gamma h \mathfrak{u}_x + g \eta _x+ \varepsilon \left(  \mathfrak{u} \mathfrak{u}_x -\frac{\gamma h^3}{3} \mathfrak{u}_{xxx}- 2 \beta h \eta_x \mathfrak{u} \right)= \mathcal{O}(\varepsilon^2).
 \end{equation}
 The energy is conserved in a sense that
 \begin{align}
     \frac{dH}{dt}&= \int_{\mathbb{R}} \left(\frac{\delta H}{\delta \mathfrak{u}(x,t)} \mathfrak{u}_t+ \frac{\delta H}{\delta \eta(x,t)} \eta_t \right)dx\nonumber \\
     &=-\left[ \frac{\delta H}{\delta \mathfrak{u}(x,t)} \frac{\delta H}{\delta \eta(x,t)}+\varepsilon^2 \gamma \frac{(\partial_x^{-1}\eta_t)^2}{2}\right]_{x=-\infty}^{x=\infty} -\varepsilon^3 2\beta \int_{\mathbb{R}} \eta_t \partial_x^{-1}(\eta_x \partial_x^{-1}\eta_t) dx \nonumber \\
     &=0-\varepsilon^3 2\beta c_0^2\int_{\mathbb{R}} [\eta_x \partial_x^{-1}(\eta \eta_x)+...] dx =- \beta c_0^2\int_{\mathbb{R}} [(\varepsilon^3 \delta) \eta^2 \eta_x +\mathcal{O}(\varepsilon^4\delta)] dx\nonumber \\
     &=-\varepsilon^3 \beta c_0^2 \left[\frac{\eta^3}{3}\right]_{x=-\infty}^{x=\infty}
 +\mathcal{O}(\varepsilon^4)=0+ \mathcal{O}(\varepsilon^4). \end{align}

The well-posedness of the Boussinesq system in the irrotational case is studied in details, see for example \cite{Bona,CGS}. In the rotational case there are new linear terms, which depend on $\gamma$ and this inevitably affects the well-posedness conditions. This problem deserves further studies.

\section{The KdV approximation} \label{kdvsec}

We will analyse now the coupled equations \eqref{eq4u1}-\eqref{eq4eta1}. It could be shown that they are asymptotically equivalent to the KdV equation \cite{KdV,ZMNP,Lan,Constantin_2011}.
Let us find the propagation speed for the travelling wave solutions of these equations in the leading order ($\varepsilon=0$), that is, solutions, which depend only on $x-c_0t.$ Then
\begin{equation}
     \begin{pmatrix} -c_0 - \gamma h & g  \\   h  & -c_0 \end{pmatrix}
      \begin{pmatrix} \mathfrak{u}' \\ \eta' \end{pmatrix} = \begin{pmatrix} 0 \\ 0 \end{pmatrix}.
\end{equation}
This homogeneous system has solution when the determinant of its matrix is zero, that is,
\begin{equation}\label{c_0}
     c_0^2+ \gamma h c_0 -gh=0.
\end{equation}
This quadratic equation has two roots (positive and negative) which describe left- and right-running waves.
Moreover, in the leading order $\mathfrak{u}=(c_0/h)\eta,$ which suggests that in the next order
\begin{equation}
    \mathfrak{u}=\frac{c_0}{h}\eta+\varepsilon (\alpha_1 \eta^2 + \alpha_2 \eta_{xx} )+\mathcal{O}(\varepsilon^2)
\end{equation}
for two yet unknown coefficients $\alpha_1$ and $\alpha_2.$  With this relation the variable $\mathfrak{u}$ can be eliminated from \eqref{eq4u1}-\eqref{eq4eta1}, but the two equations (which are in KdV form) have to be compatible, that is, identical. This condition fully determines $\alpha_1$ (from the comparison of the $\eta \eta_x$ terms) and $\alpha_2$ (from the comparison of the $ \eta_{xxx}$ terms),
\begin{equation}
\begin{split}
\alpha_1 &= - \frac{ c_0(c_0 + \gamma h  +2 \beta h^2 )}{ 2h^2(2c_0+\gamma h)},\\
\alpha_2 &=-\frac{c_0h(c_0+\gamma h)}{3(2c_0+\gamma h)} .
\end{split}
\end{equation}
And the so obtained KdV equation acquires the form
\begin{equation}\label{kdv}
    \eta_t + c_0 \eta_x + \varepsilon \frac{c_0^2 h^2}{3(2c_0+\gamma h)} \eta_{xxx}+ \varepsilon
    \frac{3g- 2\beta h c_0 + \gamma^2 h}{2c_0+\gamma h}\eta \eta_x=\mathcal{O}(\varepsilon^2).
\end{equation}
Note that $c_0$ has different values for the left- and for the right-moving waves, which are given by the roots of \eqref{c_0}. The coefficient of the nonlinear term depends on $\beta.$ In the case of vanishing $\beta$ (linear shear flow) we obtain the result from \cite{Curtin}. The KdV model is valid when all coefficients of the equation are not equal to zero. Note that the denominator $2c_0+\gamma h = \pm \sqrt{4gh +\gamma^2 h^2} \ne 0.$ The only possibility for a zero coefficient is the situation when  $3g-
2\beta h c_0 + \gamma^2 h=0,$ which requires obviously $\beta c_0 >0.$

\noindent From the formulae, discussed in the introduction, the coefficient $\beta$ usually has the form $$ \beta= \frac{n U(0)}{h^2},$$ where $n$ is a numerical factor of order 1. Let us suppose that $\gamma =0.$ Then $|c_0|=\sqrt{gh}$ and the critical condition becomes
\begin{equation}\label{crit}
     |c_0|=\sqrt{gh}=\left|\frac{2 n}{3} U(0)\right |.
\end{equation}
For ocean waves $U(0)$ is several m/s, for EUC - up to 4 m/s westwards, while $h$ is hundreds and thousands of meters and $c_0$ is much larger, hence for ocean waves such an equality is impossible. However, hypothetically, there might be other configurations in fluid mechanics where $ c_0 \sim U(0)$ and \eqref{crit} holds, then (for this critical case only) the presented analysis will require modification and different scales, as well as inclusion of higher order nonlinearities, see for example such situation in \cite{CuIv2}.

The KdV equation plays an important role in modelling in general. KdV is one of the most well known of the so-called nonlinear integrable equations.  Its solutions could be obtained by a technique, known as inverse scattering, \cite{ZMNP,Constantin_2011}. The soliton solutions, which describe stable solitary waves are of great importance, since such waves, especially in the long-wave regime are quite common and easy to observe in various situations. The one-soliton solution formula for
\begin{equation}\label{kdv1}
    \eta_t + c_0 \eta_x + b \eta_{xxx}+  a    \eta \eta_x=0
\end{equation} is
\begin{equation}
    \eta(x,t)=\frac{ 12b K^2}{a \cosh^2[K(x-x_0 - (c_0+4b K^2 )t)]},
\end{equation}
where, in our setup \begin{equation}
    a=\frac{3g-2\beta h c_0 + \gamma^2 h}{2c_0+\gamma h}, \quad b= \frac{c_0^2 h^2}{3(2c_0+\gamma h)}
\end{equation}
are the two constant coefficients, $K$  and $x_0$ are the soliton parameters, i.e. two arbitrary constants.
From here we observe only the soliton amplitude depends on $\beta$ through $a,$ while the soliton velocity $c_0+4bK^2 $ does not.

\section{Weakly nonlinear wave packets and the NLS equation}

The weakly nonlinear wave packets have the following lowest order approximation, \cite{Boyd}
\begin{equation}
\eta(x,t)=\varepsilon A(\zeta , \tau) \exp [ i (kx-\omega(k) t)]+ \text{c.c.},
\end{equation}
where $k$ is the wave-number of the carrier wave, "c.c." means complex conjugation. The complex factor $ A(\zeta , \tau)$ is called "an envelope" and evolves according to the NLS equation.
For the KdV model \eqref{kdv1}, the dispersion relation is $\omega(k)=k c_0-b k^3,$ the group velocity is
therefore $c_g(k)= \omega '(k)= c_0 - 3 b k^2.$ The "slow" space and time variable, which characterise the wave packet are $\zeta=\varepsilon (x-c_g(k)t),$ $\tau= \varepsilon^2 t.$ The NLS equation, which describes the time-evolution in the leading order of $\varepsilon$ has the form \cite{Boyd}
\begin{equation}\label{nls}
    i A_{\tau}- 3 b k A_{\zeta \zeta} + \frac{a^2}{6bk} |A|^2 A=0.
\end{equation}
The ratio of the two coefficients is $-(18 b^2 k^2)/a^2$ is always negative. This corresponds to a de-focusing NLS, which is known to possess dark solitons. This correspondence allows for an effective description of dark NLS solitons in terms of KdV solitons, see for example \cite{Hor2,Hor}.

The NLS equation \eqref{nls} is valid for long waves. In the case of finite depth and constant vorticity ($\gamma\ne 0,$ $\beta=0$) the NLS model has been derived in \cite{Kha}, where the limit to infinite depth is given as well. The ratio of the two coefficients then can have both signs, and thus both the focusing and defocusing NLS  types are possible, depending on the values of the physical parameters.

\section{Conclusions }

The constant vorticity flow satisfies automatically the vorticity equation and this is very important in the study of shear flows with linear shear. The inclusion of the quadratic term requires some further and more subtle considerations, but eventually it can be taken into account in the description of the most important propagation regime - for the long waves - where most of the energy of the wave motion is concentrated.

The extension of the approach for intermediate or short waves (deep water) is problematic, since the relative scale of the $x-$ and $z-$ derivatives is important, and it is determined by the Laplace equation for the potential $\varphi.$ In the Boussinesq approximation the $z-$ dependence of the quantities is very weak, and it sometimes is termed "columnar motion".

The presented approach for the waves with a flow is, in a sense, an alternative to the one of Johnson \cite{J} which involves the Burns condition etc. Generalisation for other types of fluid flows, for example, internal waves, is work in progress.




\small


\subsection*{Acknowledgements} The authors are thankful to the reviewers for their constructive and very helpful comments, suggestions and advice. This publication has emanated from research conducted with the financial support of Science Foundation Ireland under Grant number 21/FFP-A/9150.



\begin{thebibliography}{12}

\small

\bibitem{Bona} J.L. Bona, M. Chen, and J.-C. Saut. Boussinesq equations and other systems for small amplitude
long waves in nonlinear dispersive media. I. Derivation and linear theory, J. Non-linear Sci., 12 (2002),283--318,
\url{https://doi.org/10.1007/s00332-002-0466-4}.

\bibitem{Bow} K.F. Bowden, Physical oceanography of coastal waters. Ellis Horwood Ser. Mar. Sci. John Wiley \& Sons, Inc., Somerset, N.J. 1984, 302 p.

\bibitem{Boyd} J.P. Boyd, Dynamics of the equatorial ocean, Springer-Verlag GmbH Germany, 2018.

\bibitem{CoIv2} A. Compelli, R. Ivanov, The dynamics of flat surface internal geophysical waves with currents, J. Math. Fluid Mech. \textbf{ 19(2)}, (2017) 329--344;  	arXiv:1611.06581 [physics.flu-dyn]; \\
    \url{https://doi.org/10.1007/s00021-016-0283-4}.

\bibitem{CIMT}A.C. Compelli, R.I. Ivanov, C.I. Martin, M.D. Todorov, Surface waves over currents and uneven bottom, Deep Sea Research Part II: Topical Studies in Oceanography, \textbf{ 160}, (2019), pp 25--31; arXiv:1811.03140 [physics.flu-dyn], 
     \url{https://doi.org/10.1016/j.dsr2.2018.11.004}.

\bibitem{Constantin_2011} 
A. Constantin, \emph{Nonlinear water waves with applications to wave-current interactions and tsunamis.} \emph{CBMS-NSF Regional Conference Series in Applied Mathematics} \textbf{81} (SIAM, Philadelphia, 2011).
doi:10.1137/1.9781611971873

\bibitem{CIM-16} A. Constantin, R.I. Ivanov and C.I. Martin, Hamiltonian formulation for wave-current interactions in stratified rotational flows, {\it Archive for Rational Mechanics and Analysis}, {\bf 221} (2016) 1417--1447,  \url{https://doi.org/10.1007/s00205-016-0990-2}  (open access)

\bibitem{NearlyHamiltonian}
\newblock A. Constantin, R. Ivanov and E. Prodanov,
\newblock Nearly-Hamiltonian structure for water waves with constant vorticity,  J. Math. Fluid Mech. \textbf{ 10} (2008) no. 2, 224--237; 
 	arXiv:math-ph/0610014, 
\url{http://dx.doi.org/10.1007/s00021-006-0230-x}.

\bibitem{CJ} A. Constantin and R.S. Johnson, The dynamics of waves interacting with the Equatorial Undercurrent, Geophysical and Astrophysical Fluid Dynamics \textbf{ 109 } (2015) pp. 311--358,
\url{https://doi.org/10.1080/03091929.2015.1066785}.

\bibitem{CJ2} A. Constantin and R.S. Johnson, On the non-dimensionalisation, scaling and resulting interpretation of the classical governing equations for water waves, Journal of Nonlinear Mathematical Physics \textbf{  15}, Supplement 2 (2008), 58--73. 
\url{https://doi.org/10.2991/jnmp.2008.15.s2.5}

\bibitem{CraigGroves1}
\newblock W. Craig and M. Groves,
\newblock Hamiltonian long-wave approximations to the water-wave problem, Wave Motion \textbf{ 19} (1994), 367--389, 
\url{https://doi.org/10.1016/0165-2125(94)90003-5}

\bibitem{CGS} W. Craig, P. Guyenne, C. Sulem, The water wave problem and Hamiltonian transformation theory, In: T. Bodn\'ar et al. (eds.), {\it Waves in Flows}, Advances in Mathematical Fluid Mechanics, Springer Nature Switzerland AG 2021, 
\url{https://doi.org/10.1007/978-3-030-67845-6 4}

\bibitem{Craig1993} W. Craig and C. Sulem, Numerical simulation of gravity waves,
Journal of Computational Physics, \textbf{ 108} (1993), 73--83; \url{http://dx.doi.org/10.1006/jcph.1993.1164}.

\bibitem{CuIv} J. Cullen, R. Ivanov, On the intermediate long wave propagation for internal waves in the presence of currents, European Journal of Mechanics - B/Fluids \textbf{  84} (2020) 325--333,  	arXiv:2007.04375 [physics.flu-dyn],  
     \url{https://doi.org/10.1016/j.euromechflu.2020.07.001}

\bibitem{CuIv2} J. Cullen, R. Ivanov, Hamiltonian description of internal ocean waves with Coriolis force, Communications on Pure and Applied Analysis \textbf{  21} (2022),2291--2307;  	arXiv:2203.13940 [physics.flu-dyn], \url{https://doi.org/10.3934/cpaa.2022029} 

\bibitem{Curtin} C. Curtin and R. Ivanov, The Lagrangian formulation for wave motion with a shear current and surface tension, J. Math. Fluid Mech. \textbf{ 25}, (2023) art. 87,  	arXiv:2406.00202 [physics.flu-dyn], \url{https://doi.org/10.1007/s00021-023-00831-6}

\bibitem{Hor2} T.P. Horikis, D.J. Frantzeskakis,  Asymptotic reductions and solitons of nonlocal nonlinear Schr\"odinger equations, J. Phys. A: Math. Theor. \textbf{ 49} (2016) 205202; arXiv:1603.04714 [nlin.PS], \url{https://doi.org/10.1088/1751-8113/49/20/205202}

\bibitem{Hor} T.P. Horikis, D.J. Frantzeskakis, N.F. Smyth, Extended shallow water wave equations,
Wave Motion \textbf{ 112} (2022), 102934,  	arXiv:2205.04884 [physics.flu-dyn], \\
\url{https://doi.org/10.1016/j.wavemoti.2022.102934}.

\bibitem{Delia} D. Ionescu-Kruse, R. Ivanov, Nonlinear two-dimensional water waves with arbitrary vorticity, Journal of Differential Equations \textbf{368} (2023), 317--349;  	arXiv:2409.00446 [math.AP], \\
     \url{https://doi.org/10.1016/j.jde.2023.05.047}.

\bibitem{Iv17} R.I. Ivanov, Hamiltonian model for coupled surface and internal waves in the presence of currents, Nonlinear Analysis: Real World Applications \textbf{ 34 } (2017) 316--334;  	arXiv:1702.01441 [physics.flu-dyn], 
\url{https://doi.org/10.1016/j.nonrwa.2016.09.010}.

\bibitem{IM} R.I. Ivanov and C.I. Martin, On the time-evolution of resonant triads in rotational capillary-gravity water waves, Physics of Fluids \textbf{ 31} (2019) 117103, 	arXiv:1911.05213 [physics.flu-dyn], \url{https://doi.org/10.1063/1.5128294}.

\bibitem{J}  R.S. Johnson, A Modern Introduction to the Mathematical Theory of Water Waves (Cambridge Texts in Applied Mathematics, Series Number 19), Cambridge University Press, Cambridge, 1997.

\bibitem{KdV} D. J. Korteweg and G. De Vries, On the change of form of long waves advancing in a rectangular canal, and on a new type of long stationary waves, Philosophical Magazine, 39(240) (1895) 422-443, \url{https://doi.org/10.1080/14786449508620739}.

\bibitem{Lan} D. Lannes, The water waves problem: mathematical analysis and asymptotics (Mathematical Surveys and Monographs), American Mathematical Society, 2013.

\bibitem{Mamaev} O. I. Mamaev, Fizicheskaya okeanografiya: izbrannye trudy, (Physical Oceanography: selected papers), VNIRO, Moscow, 2000, (in Russian).

\bibitem{ZMNP} S. P. Novikov, S. V. Manakov, L. P. Pitaevsky and V. E. Zakharov, Theory of solitons: the inverse scattering method. New York: Plenum, 1984.

\bibitem{Sto} H. Stommel, Wind-drift near the equator, Deep Sea Research (1953) \textbf{ 6} (1959), pp. 298--302,
\url{https://doi.org/10.1016/0146-6313(59)90088-7}.

\bibitem{Kha} R. Thomas, C. Kharif, M. Manna,
A nonlinear Schr\"odinger equation for water waves on finite depth with constant vorticity,
Physics of Fluids \textbf{24} (2012), 127102, \\
\url{https://doi.org/10.1063/1.4768530}.


\bibitem{W} E. Wahl\'en, A Hamiltonian formulation of water waves with constant vorticity.
Lett. Math. Phys. \textbf{ 79}, (2007) 303–315, \url{ https://doi.org/10.1007/s11005-007-0143-5}.


\end{thebibliography}
\end{document}